\documentclass[aps,prb,preprint,a4paper,showpacs,showkeys,superscriptaddress]{revtex4}
\usepackage[english]{babel}
\usepackage{graphicx}
\usepackage{bm}

\begin{document}

\title{Energy threshold for nanodot creation by swift heavy ions}

\author{Marko Karlu$\rm{\check{s}i\acute{c}}$}
\affiliation{Institut $Ruder Bo\check{s}kovi\acute{c}$, P.O. Box
180, 10002 Zagreb, Croatia}
\author{Sevilay Akc\"oltekin}
\affiliation{Fakult\"at f\"ur Physik, CeNIDE, Universit\"at Duisburg-Essen,
47048 Duisburg, Germany}
\author{Orkhan Osmani}
\affiliation{Fakult\"at f\"ur Physik, CeNIDE, Universit\"at Duisburg-Essen,
47048 Duisburg, Germany}
\affiliation{University of Kaiserslautern and Research Center OPTIMAS, Physics Department, 67663 Kaiserslautern, Germany}
\author{Isabelle Monnet}
\author{Henning Lebius}
\affiliation{CIMAP (CEA-CNRS-ENSICAEN-UCBN), 14070 Caen Cedex 5,
France}
\author{Milko Jak$\rm{\check{s}i\acute{c}}$}
\affiliation{Institut $Ruder Bo\check{s}kovi\acute{c}$, P.O. Box
180, 10002 Zagreb, Croatia}
\author{Marika Schleberger}
\email{marika.schleberger@uni-due.de} 
\affiliation{Fakult\"at f\"ur Physik, CeNIDE, Universit\"at Duisburg-Essen,
47048 Duisburg, Germany}
\date{4.11.2009, V2}

\begin{abstract}
We present theoretical and experimental data on the threshold
behaviour of nanodot creation with swift heavy ions. A model
calculation based on a two-temperature model taking the spatially resolved
electron density into account gives a threshold of 12~keV/nm 
below which the energy density at the end of the track is no longer high enough to melt the material. 
In the corresponding experiments we irradiated SrTiO$_3$ surfaces under grazing
incidence with swift heavy ions. The resulting chains of nanodots
were analyzed by atomic force microscopy. In addition, samples irradiated
under normal incidence were analyzed by transmission electron microscopy.
Both experiments show two thresholds, connected to the appearance of tracks and to the 
creation of fully developed tracks, respectively. The threshold values are similar for 
surface and bulk tracks, suggesting that the same processes occur at glancing and normal 
incidence. The experimental threshold for the formation of fully developed tracks 
compares well to the value obtained by the theoretical description.
\end{abstract}


\maketitle 

\section{Introduction}

The irradiation of solid matter with heavy ions of MeV energy has
long been known to result in structural modifications ranging from
stoichiometric or geometric defects and amorphization in the bulk 
up to the creation of hillocks and chains of hillocks on the surface
\cite{Mueller1998,El-Said2004,Awazu2006,Khalfaoui2006,Akcoeltekin2007}.
In this energy range the modification of the material is not due 
to direct collisions of the projectile ion with the atoms of the 
target material (nuclear stopping power regime) but rather to a very intense interaction of the 
projectile with the electronic system of the target (electronic stopping power regime). 
How this electronic 
excitation is transformed into material modifications depends very 
much on the physical properties of the target material itself. Most 
insulating materials exhibit nanosized hillocks at the impact zone
of the ion when irradiated normal to the surface. This effect has 
been known and studied in detail for quite some time (see e.g. 
\cite{Klaumuenzer2006,Schiwietz2000,Toulemonde2006}) but is still not fully understood. 

A rather successfull approach to describe the transformation of an
electronic excitation into a heated lattice is based on 
a two-temperature model (TTM) and requires solving the coupled 
differential equations for the electron and the phonon system, 
respectively \cite{Toulemonde1992}. The model has been used to 
explain e.g.~track radii in various materials created by
irradiation perpendicular with respect to the surface under the condition
that the electron-phonon coupling constant $g$ is a fitting parameter.

Recently, the two-temperature model has been modified to include the spatially resolved 
electron density instead of a homogeneous free electron gas \cite{Akcoeltekin2007} 
which is important if the irradiation takes place under glancing angles. In this case,
a single ion is able to create a chain of hillocks along its otherwise latent track 
within the bulk. According to the model, the chains occur due to the non-homogeneous
nature of the electron density. Every time the projectile travels through a region with a high density
the energy loss is sufficiently high to feed energy into the
electronic system. This energy is transferred into the phonon system, a local melting
occurs which finally results in a nanodot on the surface. 
Since the electron density corresponds to the periodicity
of the crystal, the nanodots appear with a certain periodicity on
the surface. 

So far, these chains of hillocks have always been produced with ions 
experiencing a stopping power of about 20~keV/nm and nothing is known about the
morphology at lower stopping powers. The aim of this paper is to study ion 
induced nanodot chains in SrTiO$_3$ as a function 
of kinetic energy of the projectile in order to determine a threshold value for the stopping 
power. In addition we wish to show that the recently presented theoretical approach can be applied 
to predict a threshold value which is in agreement with experimental data. 

This paper is organized as follows: At first we will present experimental data from two different methods to 
determine the threshold energy for the production of nanodots: In section A we discuss the use 
of atomic force microscopy (AFM) to determine the minimum stopping power required to create 
a chain of nanodots on the surface. As we normally treat the surface chains as \emph{tracks on the surface}, it is interesting to compare the two manifestations of a swift-ion passage: surface chains and tracks in the bulk. Therefore, we present in section B results obtained with transmission electron microscopy 
(TEM) in order to determine the stopping power required for amorphization. We then derive a 
value for the threshold from theory and discuss the experimental data with respect to our 
theoretical predictions.

\section{Experimental determination of threshold values}
\label{Experiment}

\subsection{Grazing incidence}
\label{AFM}

The single crystal samples of
SrTiO$_3$(100) (Crystec, Berlin and Kelpin, Neuhausen) have been
irradiated without prior surface treatment at the 6.0 MV Tandem
Van de Graaff accelerator at the Institute Ruder
$\rm{Bo\check{s}kovi\acute{c}}$ in Croatia \cite{Jaksic2007}. The
irradiation was done using I ions with different kinetic
energies resulting in different stopping powers (calculated with
SRIM \cite{srim}): 6.5, 13, 18, 23, and 28~MeV yielding 2.9, 5.3,
7.2, 9.0, and 10.5~keV/nm, respectively. The angle of
incidence with respect to the surface was kept at $\phi=1.3^{\rm o}$. The
samples were oriented along the (001) direction within a few
degrees. Fluences were typically chosen to yield around 10 tracks
per $\mu$m$^2$ on average, assuming that every ion produces one
track. 
For lower stopping powers also higher fluences were used
to compensate for a possible loss in production efficiency.
Additional samples of
SrTiO$_3$(100) were irradiated under $\phi=1^{\rm o}$ at the beam lines IRRSUD (92 MeV Xe$^{23+}$, 21 keV/nm) 
and SME (700 MeV Xe$^{23+}$, 29 keV/nm) at the GANIL in Caen, France. 
After irradiation the samples were analyzed by AFM 
in the tapping mode under ambient conditions. Some samples have been cleaned prior to
AFM-measurements by snow-cleaning and wiping with ethanol. By
comparing chains of nanodots from irradiated samples with and
without cleaning treatment we made sure that this
procedure does not affect neither the shape nor height of the
hillocks.

\begin{figure}
\includegraphics[width=12cm]{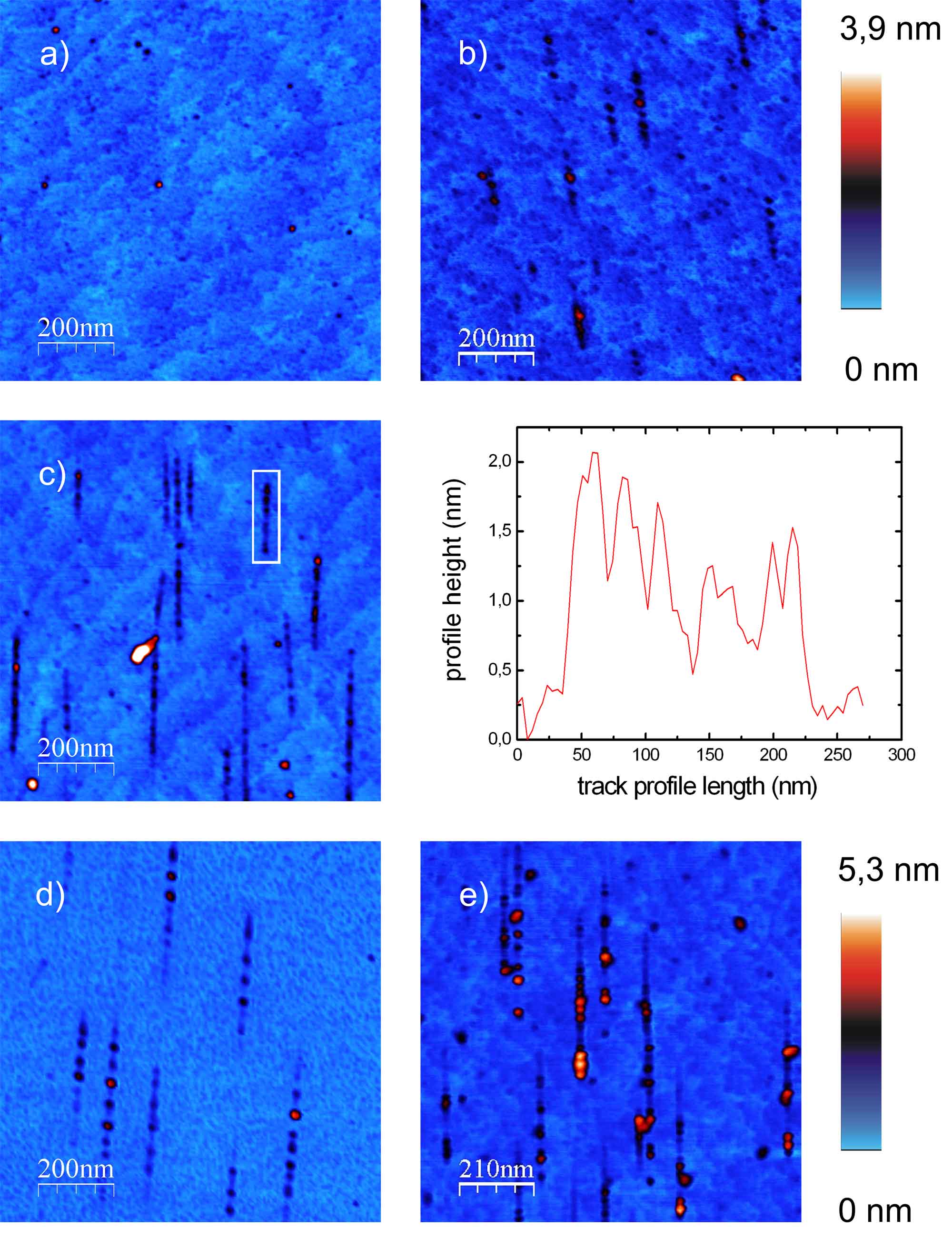}
\caption{Topography images of a SrTiO$_3$ surface irradiated with different energy losses under grazing incidence. I ions under 1.3$^{\rm o}$ were used in a)-c), Xe ions under 1$^{\rm o}$ in d) and e). Frame size: 1~$\mu$m $\times$ 1~$\mu$m. To enhance the contrast false colouring was used. Electronic energy losses in the sub-figures are 
a) 5.3~keV/nm, b) 7.2~keV/nm, c) 10.5~keV/nm with a height profile of the marked track, d) 21~keV/nm and e) 29~keV/nm, respectively.} \label{ketten}
\end{figure}

Fig.~\ref{ketten}c shows an AFM image of the typical chains that
are produced on a SrTiO$_3$ surface after irradiation with 10.5~keV/nm ions at a glancing angle of $\phi=1.3^{\rm o}$. 
Each chain in fig.~\ref{ketten}c
is produced by a single ion because 
the average number of chains ($\approx$ 10 per $\mu$m$^2$) corresponds 
well with the nominal fluence of 1x10$^9$ ions/cm$^2$. Thus, the production efficiency
is close to one. The chain length is about 300~nm and the
individual hillocks within the chains are a few nanometers high.
At this stopping power height and diameter of the dots appear to
be the same as found in earlier experiments with higher stopping
powers, see figs.~\ref{ketten}d and \ref{ketten}e. For a general discussion of the velocity effect in the electronic energy loss region on the measured chains of hillocks, see \cite{Akcoeltekin2009}.

In fig.~\ref{ketten}a and \ref{ketten}b we show a representative AFM image of
samples that have been irradiated with ions of 5.3~keV/nm and 7.2~keV/nm, respectively (same fluence, same angle as
in fig.~\ref{ketten}c). Ions with 5.3~keV/nm (fig.~\ref{ketten}a) as well as with lower energy losses produce no visible chains. First chains were detected at 7.2~keV/nm (fig.~\ref{ketten}b). As can be seen, the average chain length as well as the number of chains is smaller than for higher energy losses. Our statistic was not sufficient to quantitatively discuss the chain production efficiency (chains per incident ion), but we were able to study the length of the chains.

\begin{figure}
\includegraphics[width=8cm]{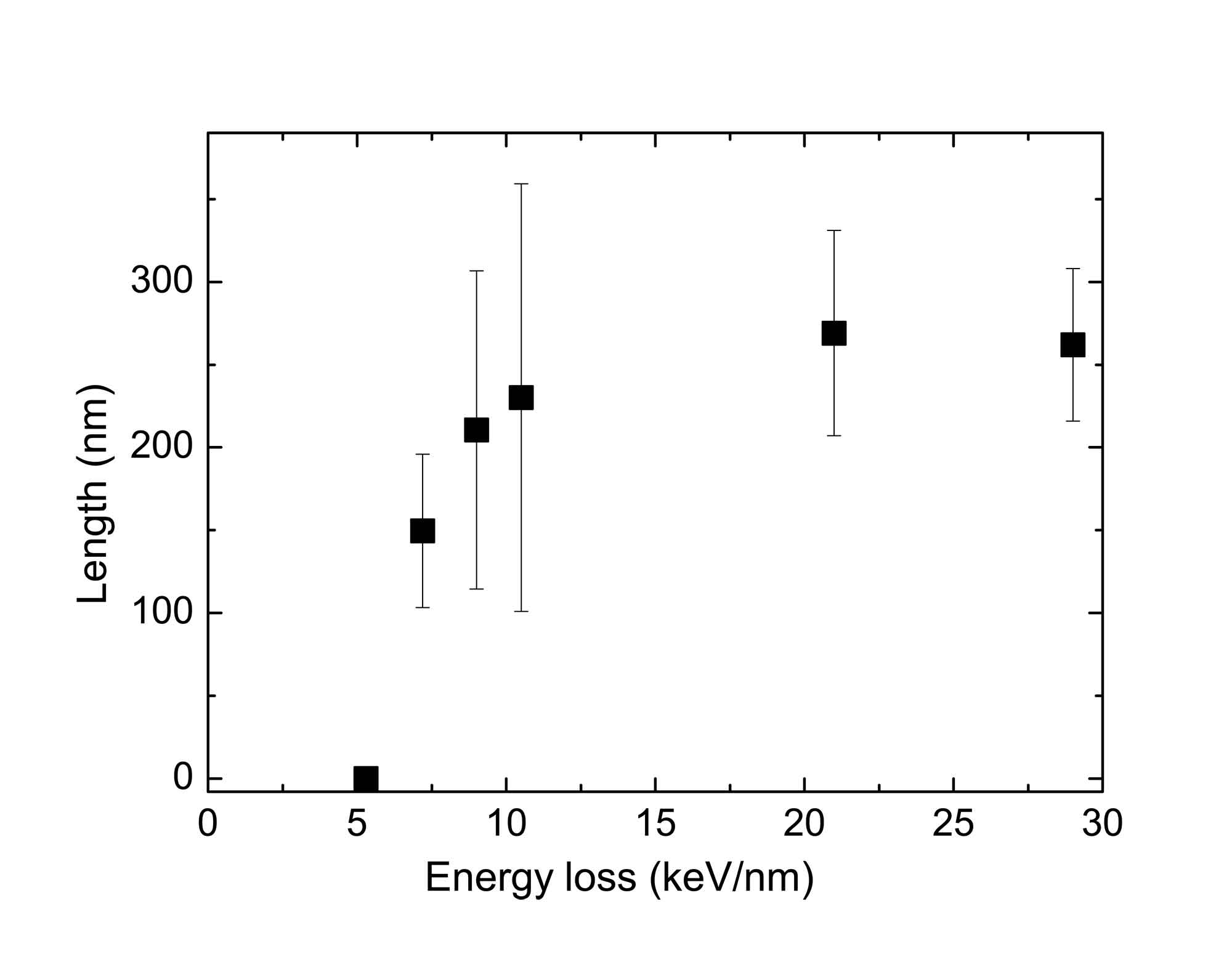}
\caption{Chain length as a function of stopping power. The data points at 21~keV/nm and at 29~keV/nm
are taken from samples irradiated with Xe ions under 1$^{\rm o}$, the chain length was corrected to present an incidence angle of 1.3$^{\rm o}$, i.e.\ they were multiplied with $\cot 1.3^{\rm o} / \cot 1^{\rm o} = 1.3$. The other data points are from irradiations with iodine ions under 1.3$^\circ$.}
\label{kettenE}
\end{figure}

To demonstrate the evolution of chains as a function of ion energy more clearly
we have performed a statistical analysis of at least 60 chains for each kinetic energy, see fig.~\ref{kettenE}. It should be noted that the error bar in the length measurement does not originate from uncertainties in the measurement itself, but from the statistical nature of the interaction process.
From fig.~\ref{kettenE} we find that the minimum stopping power to create a chain
is between 5.5~keV/nm and 7.5~keV/nm.
Up to a stopping power of between 10.5~keV/nm and 21~keV/nm the 
chain length increases with increasing kinetic energy until a constant chain length is reached. 
The SRIM calculations as well as our calculations in section III were made for Xe ions. Because iodine is right next to xenon in the periodic table, there is not much difference between those two ions with respect to the modifications they produce. Experimentally, we have seen no difference in chains created by iodine and Xe ions.

\subsection{Perpendicular incidence}
\label{TEM}

A SrTiO$_3$ single crystal (Crystec Berlin), cut along the (100)-plane, was irradiated under normal incidence with 92~MeV Xe$^{23+}$ ions with a fluence of 2$\times10^{13}$ ions/cm$^2$ at the IRRSUD beamline at GANIL, Caen, France. At these fluences, individual tracks are covering each other. After irradiation the sample was cut and glued together with epoxy, in order to produce two layers facing each other. This sandwich was then encased and glued in a 3~mm thin walled tube. The tube was cut into 1~mm thick discs. These discs were mechanically thinned to about 100~$\mu$m in thickness and dimpled so that the center thickness was less than 5~$\mu$m. Finally, a Gatan's Precision Ion Milling system was used to thin the specimen from both sides with 3.5 keV Ar$^+$ ions at 5$^\circ$ incidence angle until perforation. These targets were studied with a 2010F Jeol Transmission Electron Microscope equipped with a 200~keV field electron gun.

\begin{figure}
\includegraphics[width=12cm]{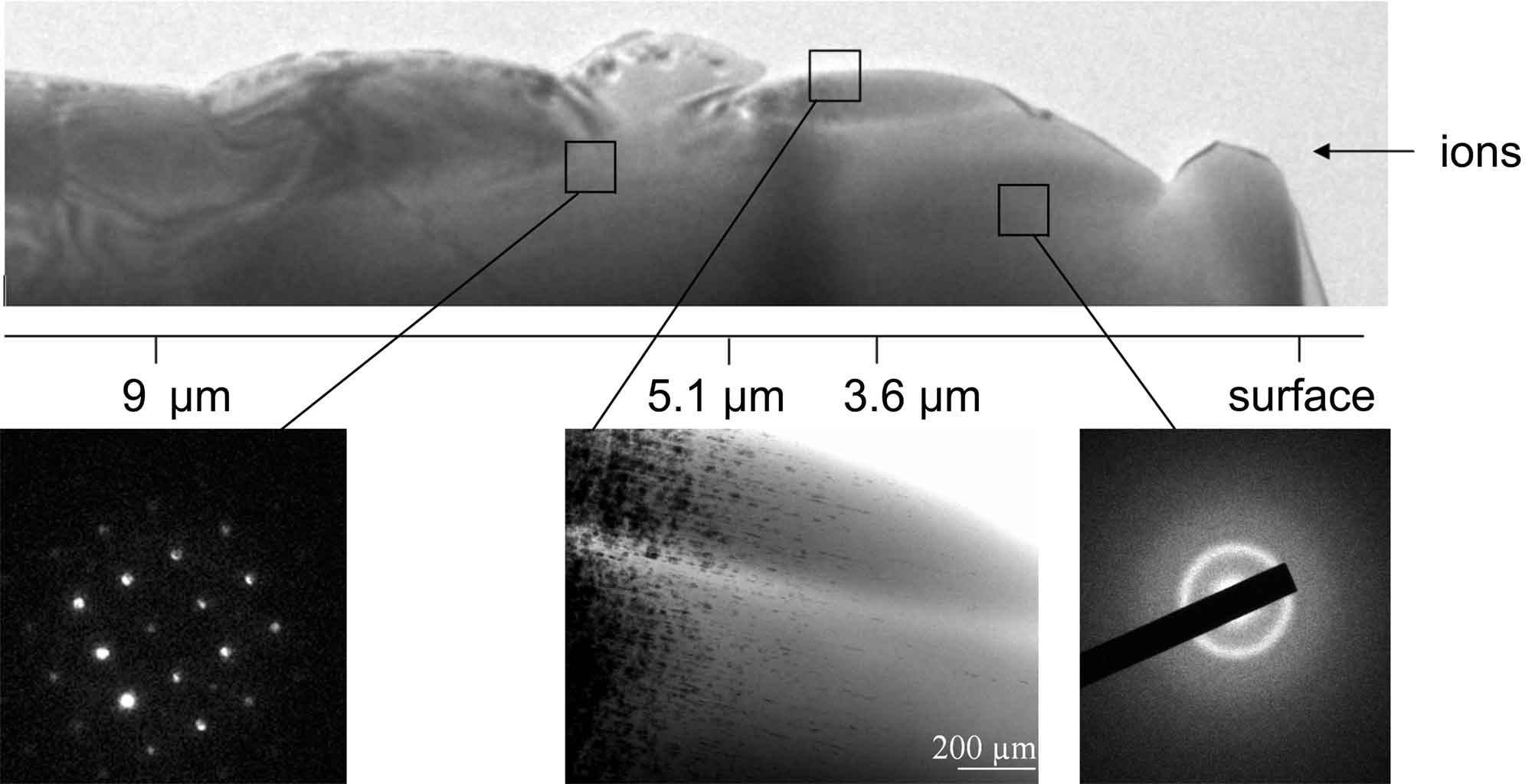}
\caption{Top part: TEM image of a SrTiO$_3$ sample irradiated under normal incidence. Ions enter on the right side, see arrow. The axis below the image shows the depth from the surface. Lower part: center image is a magnification of the central part of the image on top. The left and right images on the lower part show the diffraction patterns obtained at ion penetration depths of 6~$\mu$m and 3~$\mu$m, respectively.}
\label{tem}
\end{figure}

\begin{figure}
\includegraphics[width=8cm]{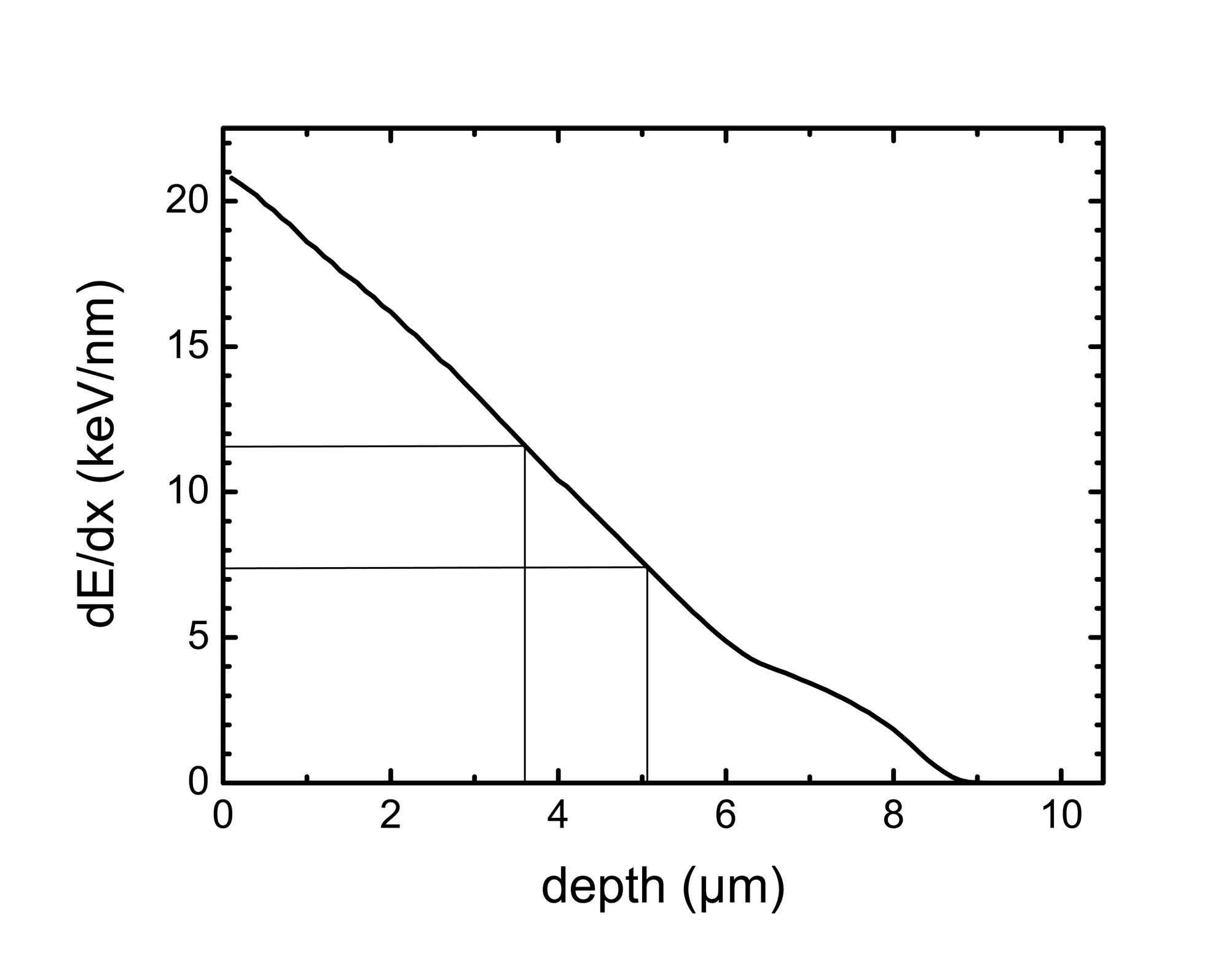}
\caption{SRIM simulation of Xe ions impinging on a SrTiO$_3$ single-crystal at 92 MeV kinetic energy. Shown is the electronic energy loss as a function of the penetration depth of the incident ion. Thin lines denote the thresholds for complete amorphization and for the appearance of partially amorphized tracks, respectively, as determined from the TEM images in fig.~3.}
\label{srim}
\end{figure}

A TEM-image of that cross section is shown in the upper part of Fig.~\ref{tem}. The bright triangle on the right side stems from the glue, denoting the surface region of the crystal. The ions penetrate the crystal in the direction indicated by the arrow. In the first part along the trajectory, the crystal was amorphized by the projectile, as can be seen from the diffraction pattern on the right side, taken from this part of the crystal. This first part extends from the surface to a penetration depth of $\simeq$ 3.6~$\mu$m. To correlate this data with the stopping power we performed SRIM calculations. The result is shown in fig.~\ref{srim}. In the first part, where the crystal was amorphized, the electronic energy loss decreases from 21~keV/nm to 11.7~keV/nm. From this we determine the threshold of amorphization to be about 11.7~keV/nm.

At larger penetration depths, partial amorphization is seen in fig.~\ref{tem}. This continues to a depth of 5.1~$\mu$m. At this depth, the energy loss is 7.3~keV/nm, according to SRIM, shown in fig.~\ref{srim}. At even larger penetration depths, the material is still mono-crystalline, as seen by the diffraction pattern on the left in fig.~\ref{tem}. Therefore we can distinguish two thresholds, one at 7.3~keV/nm for the appearance of partially amorphized tracks and one at 11.7~keV/nm for complete amorphization.

\section{The Two-Temperature model}
\label{Theorie}

After we have experimentally determined the minimum stopping power required to create a chain of nanodots we
focus on the theoretical description. Unfortunately, theoretical models describing the 
hillock formation in detail are still lacking. However, because we are only interested in 
the threshold energy a modified two-temperature model can be used. The basic procedure to determine 
the threshold energy is described in detail in \cite{Akcoeltekin2008}, and will be explained here only insofar, as is necessary to understand the additions we have made to the model. 

\begin{figure}
\includegraphics[width=8cm]{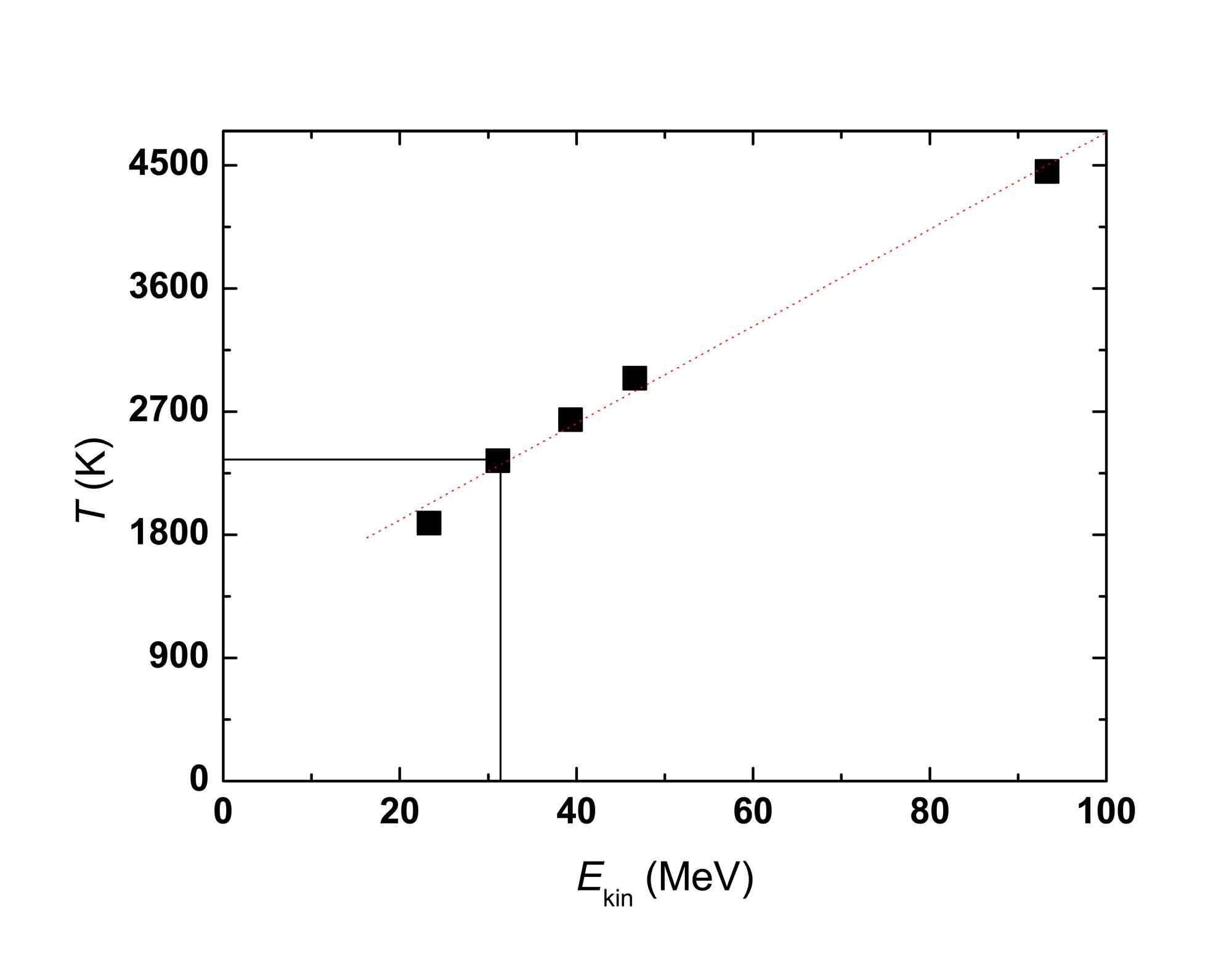}
\caption{Temperature in K at the surface of SrTiO$_3$ if irradiated by Xe ions under different kinetic energies. To reach the melting
temperature of SrTiO$_3$ the projectile has to have an energy of 31~MeV
corresponding to a stopping power of $dE/dx=$12~keV/nm. The thin
line represents a linear fit to the data. }\label{schwelle}
\end{figure}

In order to determine the minimal necessary stopping power for hillock creation the following scheme was applied. The calculations were performed on the basis of a 3D two-temperature model using space- and time-resolved electronic stopping powers for the actual case. It was proven before, that there is a fixed maximum depth for the incident ion from where it is possible to create surface hillocks due to the electronic energy loss \cite{Akcoeltekin2007,Akcoeltekin2009}. Therefore, the travelled length of the incident ion at the point perpendicular to the last hillock can easily be extracted from the incident angle and the above-mentioned maximum depth out of geometrical calculations. In the next step, we calculate the energy loss of the ion at exactly this point for different kinetic energies and use it as a source for the 3D TTM. In this way, surface temperatures at the end of the hillock chain can be obtained as a function of the incident ion energy, see fig.~\ref{schwelle}. Assuming that the melting temperature ($T_{melt}=2353$~K \cite{handbook}) has to be reached in order to induce surface modifications, the incident ion has to have a minimum kinetic energy of 31~MeV, corresponding to an energy loss of 12.7~keV/nm. This calculation ignores the necessary (but unknown) heat of fusion, and the energy loss obtained in this way therefore presents a lower limit. Additionally, this model calculation was done for Xe ions, the minimum kinetic energy value for I ions would be slightly higher to achieve the same stopping power.

\section{Discussion}
In table I, all the threshold values obtained for glancing and normal incidence, as well as the theoretical prediction are shown.

\begin{table}
\begin{tabular}[t]{ c c c }
\\\hline\hline
Grazing angle & Appearance of hillocks & Constant length of chains\\
& 5.3 - 7.2 & 10.5 - 21\\\hline
Normal incidence & Appearance of tracks & Amorphization\\
& 7.3 & 11.7\\\hline
Theory (grazing angle)& - & 12.7\\\hline\hline
\end{tabular}
\caption{Threshold values (in keV/nm) obtained by experiments and theory.}
\end{table}

Under glancing angles, tracks show up at kinetic energy losses larger than (5.3~-~7.2)~keV/nm, when studied by means of AFM (section IIA). This fits well with the TEM measurements (section IIB), where the appearance of tracks was seen for energy losses of 7.3~keV/nm.

The TEM measurements show at lower energy loss values discontinuous tracks; and the AFM measurements reveal for lower energy losses only tracks which are shorter than those with higher energy losses. Amorphization in the bulk as well as fully developed surface tracks can only be seen at energy losses above 11.7~keV/nm and (10.5~-~21)~keV/nm, respectively.

This second threshold fits well with the calculated threshold of 12.7~keV/nm, where melting of the surface over the full surface track length was defined as a requirement in order to produce hillocks, see section III. As was shown before \cite{Akcoeltekin2008} the track length for high energy losses is given by a maximum depth $d$, from which surface modifications are possible. The depth is for SrTiO$_3$ $d\approx 9$~nm, and the track length $l$ can then be calculated by $l = d \cot \phi$. For sufficiently high energy losses, $d$ is constant even if the initial energy loss of the impinging ion changes by over 70\% \cite{Akcoeltekin2009}. If the energy loss approaches the threshold, the remaining energy loss at the maximum depth $d$ may not be enough for surface modification. This can explain, why we observe increasing surface track lengths with increasing energy loss in fig. 2.

The theoretically predicted threshold assumes that the total chain length is constant in order to make assumptions on the necessary stopping power. The calculation of the lower threshold is in principle possible as well, but this would require an additional variation of the surface track length which would tremendously increase the computational effort. The discovery of two different values for the hillock creation is not a contradiction, it merely reflects that the hillock creation at stopping powers $\geq 12.7$~keV/nm are independant of statistical fluctuations. Below this value the individual hillock creation is a more statistical  phenomenon which results in "incomplete" tracks.

The determination of the threshold with chains has two main advantages. Experiments performed under perpendicular incidence create single hillocks which are difficult to identify in \emph{ex situ} experiments, especially if the production efficiency goes down or the single hillocks become smaller when approaching the threshold. A chain is easily spotted and can be unambiguously identified as an ion-induced feature, for example the direction must comply with the direction of the irradiation. Since we made no specific assumptions with respect to the material, we believe that the method is capable of predicting the minimal energy required to create nanodots in many other insulating materials of technological importance, such as Al$_2$O$_3$ or TiO$_2$. The determination of the threshold energy by means of analyzing surface tracks has proven useful even for systems like SiO$_2$ \cite{Carvalho2007} or III-V semiconductors \cite{Mansouri2008} where the connection to the electron density might not be as straightforward.

\section*{Acknowledgement}
We thank R. Meyer for calculating the electron density and for
stimulating discussions. Financial support by the DFG - SFB 616:
{\it Energy dissipation at surfaces} and by the International
Atomic Energy Agency (IAEA) research project CRO12925 is
gratefully acknowledged. Experiments were performed at the Institute Ruder $\rm{Bo\check{s}kovi\acute{c}}$ in Croatia and GANIL, France.

\bibliography{Threshold}

\begin{thebibliography}{10}

\bibitem{Mueller1998}
A.~M\"uller, R.~Neumann, K.~Schwartz, and C.~Trautmann.
\newblock {Scanning force microscopy of heavy-ion tracks in lithium fluoride}.
\newblock {\em {Nucl.\ Instr.\ and Meth.\ B}}, {146}:{393}, {1998}.

\bibitem{El-Said2004}
A.~S. El-Said, M.~Cranney, N.~Ishikawa, A.~Iwase, R.~Neumann, K.~Schwartz,
  M.~Toulemonde, and C.~Trautmann.
\newblock {Study of heavy-ion induced modifications in BaF$_2$ and LaF$_3$
  single crystals}.
\newblock {\em {Nucl.\ Instr.\ and Meth.\ B}}, {218}:{492}, {2004}.

\bibitem{Awazu2006}
K.~Awazu, X.~Wang, M.~Fujimaki, T.~Komatsubara, T.~Ikeda, and Y.~Ohki.
\newblock {Structure of latent tracks in rutile single crystal of titanium
  dioxide induced by swift heavy ions}.
\newblock {\em {Journal of Applied Physics}}, {100}:{044308}, {2006}.

\bibitem{Khalfaoui2006}
N.~Khalfaoui, M.~G\"orlich, C.~M\"uller, M.~Schleberger, and H.~Lebius.
\newblock {Latent tracks in CaF$_2$ studied with atomic force microscopy in air
  and in vacuum}.
\newblock {\em {Nucl.\ Instr.\ and Meth.\ B}}, {245}:{246}, {2006}.

\bibitem{Akcoeltekin2007}
E.~Akc\"oltekin, T.~Peters, R.~Meyer, A.~Duvenbeck, M.~Klusmann, I.~Monnet,
  H.~Lebius, and M.~Schleberger.
\newblock {Creation of multiple nanodots by single ions}.
\newblock {\em {Nature Nanotechnology}}, {2}:{290}, {2007}.

\bibitem{Klaumuenzer2006}
S.~Klaum\"unzer.
\newblock {Thermal-spike Models for Ion Track Physics: A Critical Examination}.
\newblock {\em {Mat.\ Fys.\ Medd.\ Dan.\ Vid.\ Selsk.}}, {52}:{293}, {2006}.

\bibitem{Schiwietz2000}
G.~Schiwietz, G.~Xiao, E.~Luderer, and P.~L. Grande.
\newblock {Auger electrons from ion tracks}.
\newblock {\em {Nucl.\ Instr.\ and Meth.\ B}}, {164}:{353}, {2000}.

\bibitem{Toulemonde2006}
M.~Toulemonde, C.~Dufour, and E.~Paumier.
\newblock {The ion-matter interaction with swift heavy ions in the light of
  inelastic thermal spike model}.
\newblock {\em {Acta Phys.\ Pol.\ A}}, {109}:{311}, {2006}.

\bibitem{Toulemonde1992}
M.~Toulemonde, C.~Dufour, and E.~Paumier.
\newblock {Transient thermal-process after a high-energy heavy-ion irradiation
  of amorphous metals and semiconductors}.
\newblock {\em {Phys.\ Rev.\ B}}, {46}:{14362}, {1992}.

\bibitem{Jaksic2007}
M.~Jaksic, I.~Bogdanovic Radovic, M.~Bogovac, V.~Desnica, S.~Fazinic,
  M.~Karlusic, Z.~Medunic, H.~Muto, Z.~Pastuovic, Z.~Siketic, N.~Skukan, and
  T.~Tadic.
\newblock {New capabilities of the Zagreb ion microbeam system}.
\newblock {\em {Nucl.\ Instr.\ and Meth.\ B}}, {260}:{114}, {2007}.

\bibitem{srim}
J.~F. Ziegler and J.~P. Biersack.
\newblock {The stopping and range of ions in matter (SRIM)}.
\newblock {\em {http://www.srim.org}}, {2008}.

\bibitem{Akcoeltekin2009}
S.~Akc\"oltekin, E.~Akc\"oltekin, M.~Schleberger, and H.~Lebius.
\newblock {Scanning probe microscopy investigation of nanostructured surfaces
  induced by swift heavy ions}.
\newblock {\em {J.\ Vac.\ Science Techn.\ B}}, {27}:{944}, {2009}.

\bibitem{Akcoeltekin2008}
E.~Akc\"oltekin, S.~Akc\"oltekin, O.~Osmani, A.~Duvenbeck, H.~Lebius, and
  M.~Schleberger.
\newblock {Swift heavy ion irradiation of SrTiO3 under grazing incidence}.
\newblock {\em {New J.\ Phys.}}, {10}:{053007}, {2008}.

\bibitem{handbook}
H.~Dittrich, N.~Karl, S.~K\"uck, H.~W. Schock, and O.~Madelung.
\newblock {\em {Semiconductors: Ternary Compounds, Organic Semiconductors}},
  volume {41E} of {\em {Landolt-B\"ornstein - Group III Condensed Matter}}.
\newblock {Springer}, {2000}.

\bibitem{Carvalho2007}
A.~M. J.~F. Carvalho, M.~Marinoni, A.~D. Touboul, C.~Guasch, H.~Lebius,
  M.~Ramonda, J.~Bonnet, and F.~Saigne.
\newblock {Discontinuous ion tracks on silicon oxide on silicon surfaces after
  grazing-angle heavy ion irradiation}.
\newblock {\em {Appl.\ Phys.\ Lett.}}, {90}:{073116}, {2007}.

\bibitem{Mansouri2008}
S.~Mansouri, P.~Marie, C.~Dufour, G.~Nouet, I.~Monnet, and H.~Lebius.
\newblock {Swift heavy ions effects in III-V nitrides}.
\newblock {\em {Nucl.\ Instr.\ and Meth.\ B}}, {266}:{2814}, {2008}.

\end{thebibliography}
\bibliographystyle{unsrt}

\end{document}